\documentclass[10pt]{blois07}
\usepackage{graphicx}
\usepackage{cite,./mcite}

\begin{document}
\title{Angular decorrelations in Mueller-Navelet jets and DIS}
\author{Agust{\' \i}n~Sabio~Vera$^1$\protect\footnote{~~talk presented 
at EDS07},
Florian~Schwennsen$^2$}
\institute{$^1$ Physics Department, 
Theory Division, CERN,  CH--1211 Geneva 23, Switzerland,\\ 
$^2$ II. Institut f\"{u}r Theoretische Physik, 
Universit\"{a}t Hamburg, Luruper Chaussee 149 D--22761 Hamburg, Germany.}
\maketitle
\begin{abstract}
We discuss the azimuthal angle decorrelation of Mueller--Navelet jets at 
hadron colliders and forward jets in Deep Inelastic Scattering within the BFKL 
framework with a NLO kernel. We stress the need of collinear improvements to 
obtain good perturbative convergence. We provide estimates of these 
decorrelations for large rapidity differences at the Tevatron, LHC and HERA.
\end{abstract}

\section{BFKL cross sections}
In this contribution we discuss the results recently obtained 
in~\cite{Vera:2006un,*Vera:2007kn,*Schwennsen:2007hs,*Vera:2007dr} where 
azimuthal angle correlations in Mueller--Navelet jets~\cite{Mueller:1986ey} 
and forward jets at HERA using the Balitsky--Fadin--Kuraev--Lipatov (BFKL) 
equation in the next--to--leading (NLO) 
approximation~\cite{Fadin:1998py,*Ciafaloni:1998gs} were investigated 
(related works can be found 
in~\cite{Marquet:2005vp,*Kepka:2006cg,*Kepka:2006xe,*Marquet:2007xx}). 
In this section we focus on normalized differential cross sections for 
Mueller--Navelet jets, which are quite insensitive to parton distribution 
functions. This justifies the use of partonic cross sections which can 
be written as
\begin{eqnarray}
\frac{d {\hat \sigma}}{d^2\vec{q}_1 d^2\vec{q}_2} &=& 
\frac{\pi^2 {\bar \alpha}_s^2}{2} 
\frac{1}{q_1^2 q_2^2} \int \frac{d\omega}{2 \pi i} e^{\omega {\rm Y}} 
f_\omega \left(\vec{q}_1,\vec{q}_2\right),
\end{eqnarray}
where ${\bar \alpha}_s = \alpha_s N_c / \pi$, $\vec{q}_{1,2}$ are the 
transverse momenta of the tagged jets, and Y their relative rapidity. The 
Green's function carries the bulk of the Y dependence and is the solution to 
the NLO BFKL equation,
\begin{eqnarray} 
\left(\omega - {\bar \alpha}_s {\hat K}_0 
- {\bar \alpha}_s^2 {\hat K}_1\right) {\hat f}_\omega =  {\hat 1},
\end{eqnarray}
which acts on the basis including the azimuthal angle, {\it i.e.},
\begin{eqnarray}
\left< \vec{q}\right|\left.\nu,n\right> = \frac{1}{\pi \sqrt{2}} 
\left(q^2\right)^{i \nu -\frac{1}{2}} \, e^{i n \theta}.
\end{eqnarray} 
As Y increases the azimuthal dependence is mainly driven by the kernel and 
therefore we use LO jet vertices which are simpler than the NLO 
ones~\cite{Bartels:2001ge,*Bartels:2002yj}. The 
differential cross section in the azimuthal angle $\phi=\theta_1-\theta_2-\pi$,
 with $\theta_i$ being the angles of the two tagged jets, reads
\begin{eqnarray}
\frac{d {\hat \sigma}\left(\alpha_s, {\rm Y},p_{1,2}^2\right)}{d \phi}  &=&
\frac{\pi^2 {\bar \alpha}_s^2}{4 \sqrt{p_1^2 p_2^2}} \sum_{n=-\infty}^\infty 
e^{i n \phi} \, {\cal C}_n \left({\rm Y}\right), 
\end{eqnarray}
where
\begin{eqnarray}
{\cal C}_n \left({\rm Y}\right) &=&
\frac{1}{2 \pi}\int_{-\infty}^\infty 
\frac{d \nu}{\left(\frac{1}{4}+\nu^2\right)}
\left(\frac{p_1^2}{p_2^2}\right)^{i \nu} 
e^{\chi \left(\left|n\right|,\frac{1}{2}+ i \nu,{\bar \alpha}_s 
\left(p_1 p_2\right)\right){\rm Y} },
\end{eqnarray}
and the NLO kernel can be written as
\begin{eqnarray}
\chi \left(n,\gamma,{\bar \alpha}_s\right) &=& 
{\bar \alpha}_s \chi_0\left(n,\gamma\right)
+{\bar \alpha}_s^2 \left(\chi_1\left(n,\gamma\right)
-\frac{\beta_0}{8 N_c} \frac{\chi_0\left(n,\gamma\right)}
{\gamma \left(1-\gamma\right)}\right).
\end{eqnarray}
The eigenvalue of the LO kernel is $\chi_0 \left(n,\gamma\right) = 
2 \psi \left(1\right) - \psi \left(\gamma+ \frac{n}{2}\right) 
- \psi\left(1-\gamma+\frac{n}{2}\right)$,
with $\psi$ the logarithmic derivative of the Euler function. The action of 
$\hat{K}_1$, in $\overline{\rm MS}$ scheme, can be found 
in~\cite{Kotikov:2000pm}. The full cross section only depends on the $n=0$ 
component,
\begin{eqnarray}
{\hat \sigma} =  \frac{\pi^3 {\bar \alpha}_s^2}{2 \sqrt{p_1^2 p_2^2}} \, 
{\cal C}_0 \left({\rm Y}\right).
\end{eqnarray}
The average of the cosine of the 
azimuthal angle times an integer projects out the contribution from each of 
these angular components:
\begin{eqnarray}
\frac{\left<\cos{\left( m \phi \right)}\right>}{\left<
\cos{\left( n \phi \right)}\right>} &=& \frac{{\cal C}_m 
\left({\rm Y}\right)}{{\cal C}_n\left({\rm Y}\right)}
\label{Ratiosformula}
\end{eqnarray}
The normalized differential cross section is
\begin{eqnarray}
\frac{1}{{\hat \sigma}}\frac{d{\hat \sigma}}{d \phi} ~=~
\frac{1}{2 \pi}\sum_{n=-\infty}^\infty 
e^{i n \phi} 
\frac{{\cal C}_n\left({\rm Y}\right)}
     {{\cal C}_0\left({\rm Y}\right)} 
~=~ \frac{1}{2\pi}
\left\{1+2 \sum_{n=1}^\infty \cos{\left(n \phi\right)}
\left<\cos{\left( n \phi \right)}\right>\right\}.
\label{fullangular}
\end{eqnarray}
The BFKL resummation is not stable at 
NLO for zero conformal 
spin. A manifestation of this lack of convergence is what we found in 
the gluon--bremsstrahlung scheme where our NLO distributions have an 
unphysical behavior whenever the $n=0$ conformal spin appears in the 
calculation. To solve this problem we imposed compatibility with 
renormalization group evolution in the DIS limit 
following~\cite{Salam:1998tj,*Ciafaloni:2003rd,*Vera:2005jt} for all conformal 
spins. The new kernel with improvements to all orders 
reads~\cite{Vera:2006un,*Vera:2007kn,*Schwennsen:2007hs,*Vera:2007dr}
\begin{eqnarray}
\omega  &=& {\bar \alpha}_s \left(1+ {\cal A}_n {\bar \alpha}_s\right)
\left\{2 \, \psi \left(1\right) 
- \psi \left(\gamma + \frac{\left|n\right|}{2}+\frac{\omega}{2}+{\cal B}_n {\bar \alpha}_s \right) \right. \nonumber\\
&&\hspace{-1cm}- \left.\psi \left(1-\gamma + \frac{\left|n\right|}{2}+\frac{\omega}{2}+{\cal B}_n {\bar \alpha}_s \right) \right\} + {\bar \alpha}_s^2 \, \Bigg\{\chi_1 \left(\left|n\right|,\gamma\right)-\frac{\beta_0}{8 N_c} \frac{\chi_0\left(n,\gamma\right)}{\gamma \left(1-\gamma\right)}\nonumber\\
&&\hspace{-1cm}-{\cal A}_n \chi_0\left(\left|n\right|,\gamma\right)\Bigg)
+ \left(\psi'\left(\gamma + \frac{\left|n\right|}{2}\right)
+\psi' \left(1-\gamma + \frac{\left|n\right|}{2}\right) \right)
\left(\frac{\chi_0\left(\left|n\right|,\gamma\right)}{2}+{\cal B}_n\right)
\Bigg\},
\end{eqnarray}
where ${\cal A}_n$ and ${\cal B}_n$ are collinear 
coefficients~\cite{Vera:2006un,*Vera:2007kn,*Schwennsen:2007hs,*Vera:2007dr}. 
After this 
collinear resummation our observables have a good physical behavior and are 
independent of the renormalization scheme. It is very important to remark that 
the asymptotic behavior of the BFKL resummation is convergent for non zero 
conformal spins. In this sense the ideal distributions to investigate 
experimentally are those of the form $\left<cos{(m \phi)}\right> / 
\left<cos{(n \phi)}\right>$ with $m,n \neq 0$, we will see that in this case 
the difference between the LO and higher orders results is small.

\section{Phenomenology for Mueller--Navelet jets} 

The D$\emptyset$~\cite{Abachi:1996et} collaboration analyzed data for 
Mueller--Navelet jets at $\sqrt{s} = 630$ and 
1800 GeV. For the angular correlation LO BFKL predictions were first 
obtained in~\cite{DelDuca:1993mn,*Stirling:1994zs} and failed to 
describe the data estimating too much decorrelation. An exact fixed NLO 
analysis 
using JETRAD underestimated the decorrelation, while HERWIG was in agreement 
with the data.
\begin{figure}[htbp]
  \centering
\includegraphics[width=8cm]{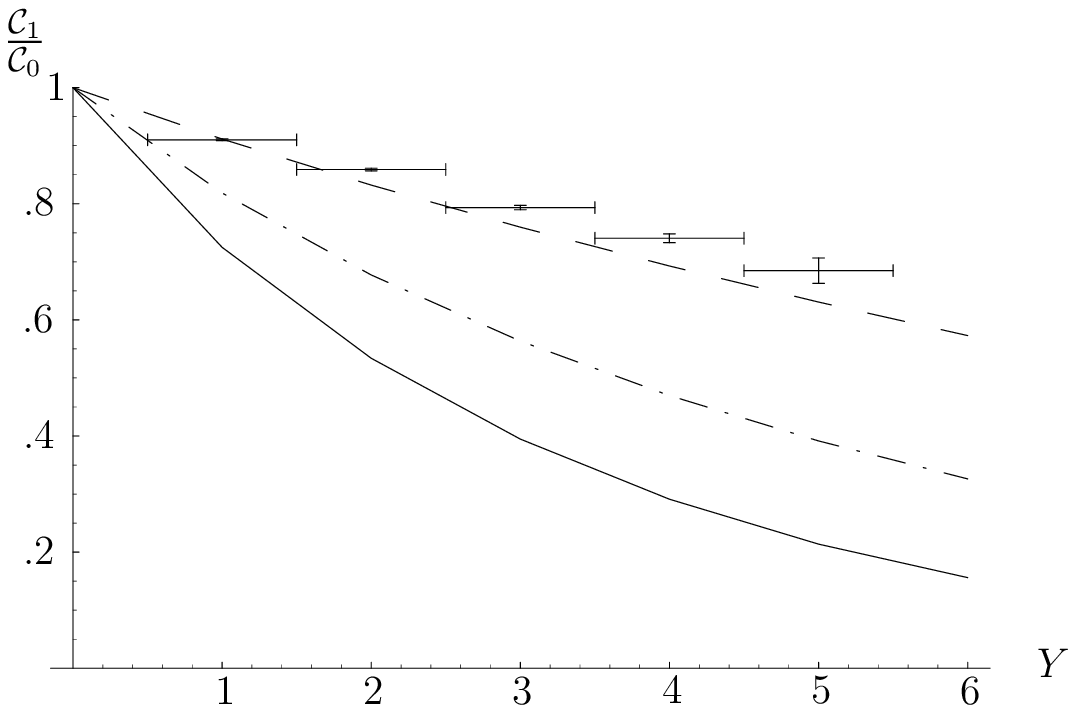}

\includegraphics[width=8cm]{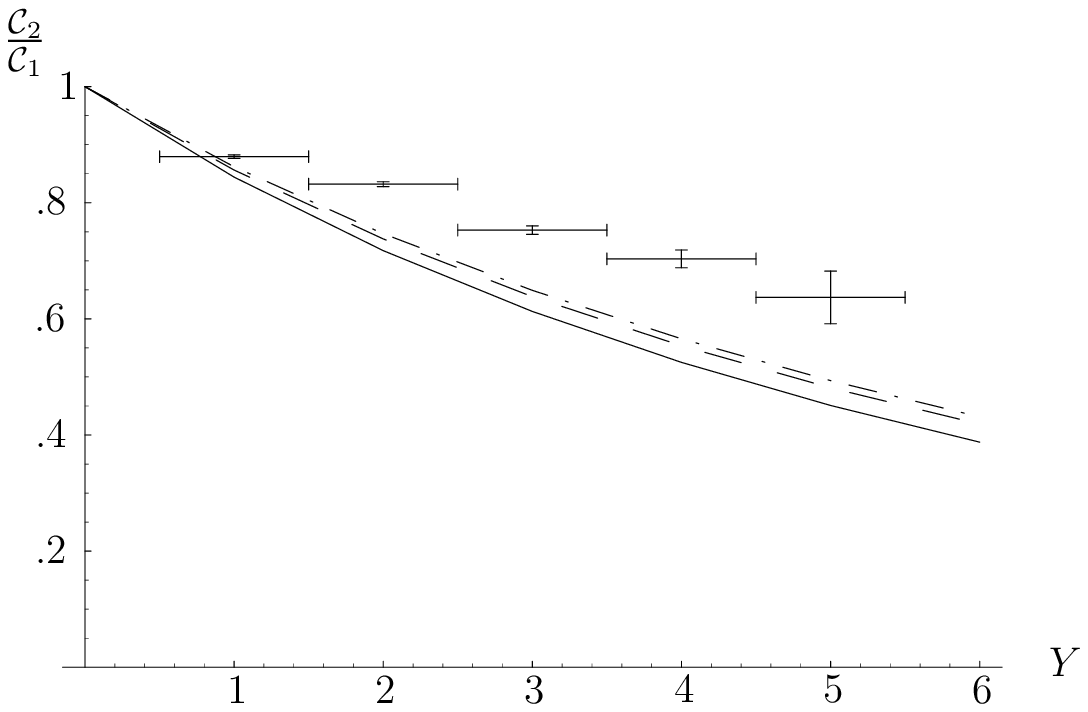}
\caption{Top: $\left<\cos\phi\right> = {\cal C}_1/{\cal C}_0$ and Bottom: 
$\frac{<\cos2\phi>}{<\cos\phi>} = \frac{{\cal C}_2}{{\cal C}_1}$, at a 
$p\bar{p}$ collider with $\sqrt{s}$ = 1.8 TeV for BFKL at LO (solid) and 
NLO (dashed). The results from the resummation presented in the text are 
shown as well (dash--dotted).}
\label{fig:tevatron1}
\end{figure}

In Fig.~\ref{fig:tevatron1} we compare the Tevatron data for 
$\left<\cos\phi\right> = {\cal C}_1/{\cal C}_0$ with our LO, NLO and resummed 
predictions. For Tevatron's cuts, where the transverse momentum for one 
jet is 20 GeV and for the other 50 GeV, the NLO calculation is instable 
under renormalization 
scheme changes. The convergence of our observables is poor 
whenever the coefficient associated to zero conformal spin, ${\cal C}_0$, is 
involved. If we eliminate this coefficient by calculating the ratios 
defined in 
Eq.~(\ref{Ratiosformula}) then the predictions are very stable, 
see Fig.~\ref{fig:tevatron1}.

\begin{figure}[htbp]
  \centering
\includegraphics[width=8cm]{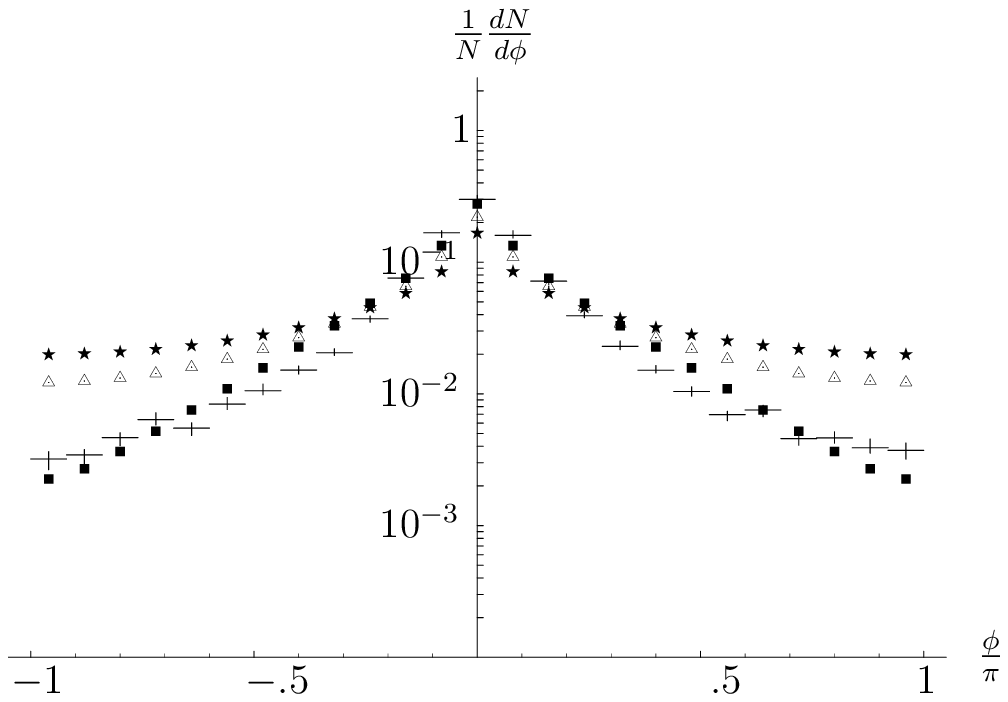}

\includegraphics[width=8cm]{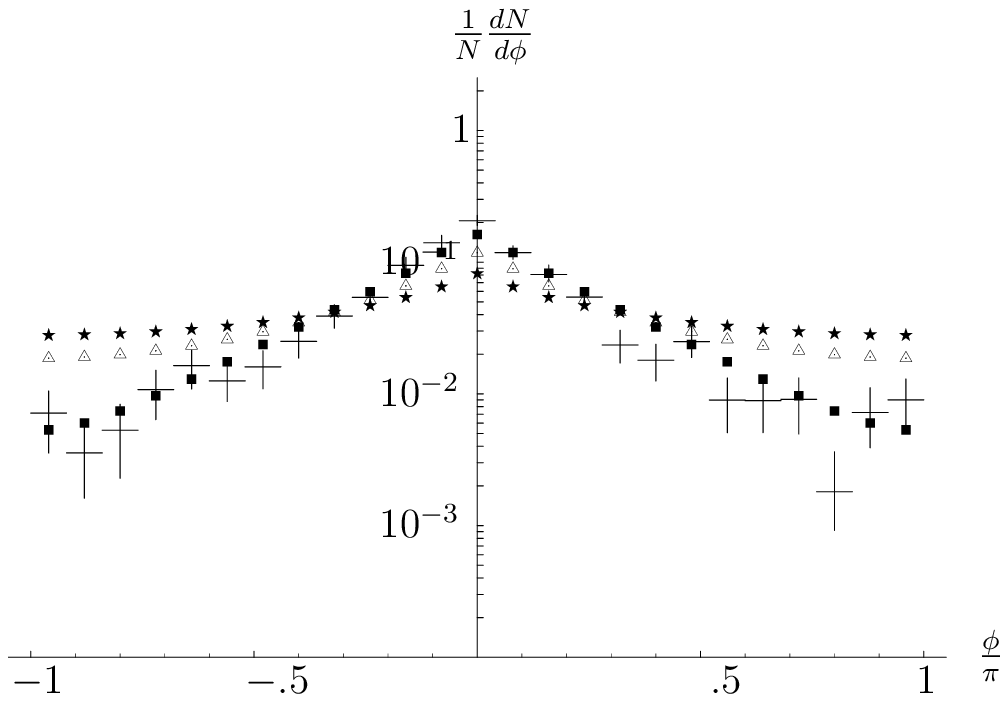}
   \caption{$\frac{1}{N}\frac{dN}{d\phi}$ in a $p\bar{p}$ collider at 
$\sqrt{s}$=1.8 TeV using a LO (stars), NLO (squares) and resummed (triangles) 
BFKL kernel. Plots are shown for Y = 3 (top) and Y = 5 (bottom).}\nonumber
  \label{fig:tevatrondsigma}
\end{figure}

\begin{figure}
\centerline{\includegraphics[width=0.5\columnwidth]{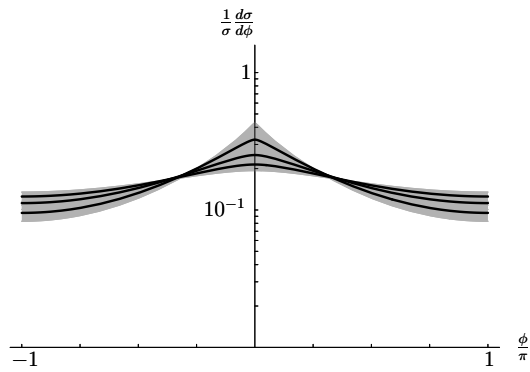}}
\caption{$\frac{1}{\sigma}\frac{d\sigma}{d\phi}$ in our resummation scheme for 
rapidities Y = 7, 9, 11 from top to bottom. The gray band reflects the 
uncertainty in $s_0$ and in the renormalization scale $\mu$.}
\label{fig:lhcdsigma}
\end{figure}

The full angular dependence studied at the Tevatron 
by the {D$\emptyset$} collaboration was published 
in~\cite{Abachi:1996et}. In Fig.~\ref{fig:tevatrondsigma} we compare this 
measurement with the predictions obtained in our approach. 
For the differential cross section we also make predictions for the LHC at larger Y in  
Fig.~\ref{fig:lhcdsigma}. 
We estimated several uncertainties in our approach which are represented by 
gray bands.

\section{Forward jets at HERA}
 
In this section we apply the BFKL formalism to predict the decorrelation in 
azimuthal angle between the electron and a forward jet associated to the 
proton in Deep Inelastic Scattering (DIS). When the separation in rapidity 
space between the scattered electron and the forward jet is large and 
the transverse momentum of the jet is similar to the virtuality of the photon 
resolving the hadron, then the dominant terms are of BFKL type. This process 
is similar to that of Mueller--Navelet jets, the only difference being 
the substitution of one jet vertex by the electron vertex describing the 
coupling of the electron to the BFKL gluon Green's function via a 
quark--antiquark pair. Azimuthal angles in forward jets were studied at LO 
in~\cite{Bartels:1996wx}. We improved their calculation by considering 
the NLO BFKL kernel. 

In the production of a forward  jet in DIS it is necessary to extract 
a parton with a large longitudinal momentum fraction $x_{\rm FJ}$ from the 
proton. When the jet is characterized by a hard scale it is possible to use 
conventional collinear factorization to describe the process and the jet 
production rate may be written as
\begin{eqnarray}
  \sigma(s) &=& \int dx_{\rm FJ}\;f_{\rm eff}(x_{\rm FJ},\mu_F^2)
\hat\sigma(\hat{s}), 
\end{eqnarray}
with $\hat\sigma(\hat{s})$ denoting the partonic cross section, and the 
effective parton density~\cite{Combridge:1983jn} being 
\begin{eqnarray}
  \label{eq:feff}
  f_{\rm eff}(x,\mu_F^2) &=& G(x,\mu_F^2)+\frac{4}{9}\sum_f\left[Q_f(x,\mu_F^2)+\bar{Q}_f(x,\mu_F^2)\right],
\end{eqnarray}
where the sum runs over all quark flavors, and $\mu_F$ stands for the 
factorization scale.

The final expression for the cross section at hadronic level is of the form
\begin{eqnarray}
  \label{eq:disfinal}
  \frac{d\sigma}{dY\;d\phi} &=& C_0(Y)+C_2(Y)\cos 2\phi, 
\end{eqnarray}
with
\begin{eqnarray}
  C_n(Y) &=& \frac{\pi^2\bar{\alpha}_s^2}{2}\int_{\rm cuts} \hspace{-.3cm}dx_{\rm FJ}\,dQ^2\,dy\,f_{\rm eff}(x_{\rm FJ},Q^2) B^{(n)}(y,Q^2,Y) \delta\left(x_{\rm FJ}-\frac{Q^2 e^Y}{ys}\right),
\end{eqnarray}
where the index in the integral sign refers to the cuts 
\begin{eqnarray}
&&20 ~{\rm GeV}^2  < Q^2   < 100 ~{\rm GeV}^2, \, \, \, \, 0.05 < y < 0.7 , 
\, \, \, \, 5\cdot 10^{-3} > x_{\rm Bj} > 4\cdot 10^{-4}.
\label{eq:heracuts}
\end{eqnarray}
The integration over the longitudinal momentum fraction $x_{\rm FJ}$ of the 
forward jet involves a delta function fixing the rapidity 
$Y=\ln x_{\rm FJ}/x_{\rm Bj}$ and $B^{(n)}$ is a complicated function 
which can be found 
in~\cite{Vera:2006un,*Vera:2007kn,*Schwennsen:2007hs,*Vera:2007dr}.

\begin{figure}[htbp]
  \centering
  \includegraphics[width=8cm]{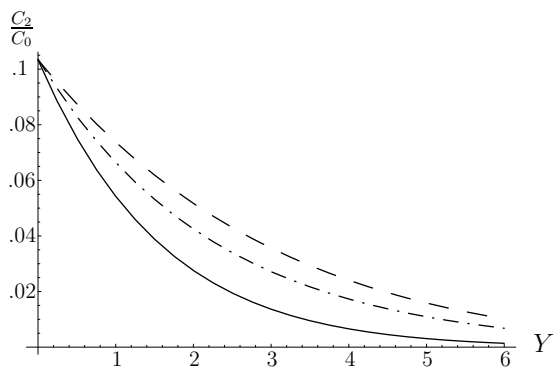}
  \caption{$<\cos 2\phi>$ at the $ep$ collider HERA at leading (solid), next to leading order (dashed), and for resummed kernel (dash-dotted). }
  \label{fig:hera1}
\end{figure}

Since the structure of the electron vertex singles out the components 
with conformal spin 0 and 2, the number of observables related to the 
azimuthal angle dependence is limited when compared to the 
Mueller--Navelet case. The most relevant observable is the dependence 
of the average $<\cos 2\phi> = C_2/C_0$ with the rapidity difference 
between the forward jet and outgoing lepton. It is natural to expect 
that the forward jet will be more decorrelated from the leptonic system 
as the rapidity difference is larger since the phase space for further 
gluon emission opens up. This is indeed what we observe in our 
numerical results shown in Fig.~\ref{fig:hera1}. We find similar results 
to the Mueller--Navelet jets case where the most reliable calculation 
is that with a collinearly--improved kernel. The main effect of the 
higher order corrections is to increase the azimuthal angle correlation 
for a given rapidity difference, while keeping the decrease of the 
correlation as $Y$ grows.

\begin{footnotesize}
\bibliographystyle{blois07} 
{\raggedright
\bibliography{blois07}
}
\end{footnotesize}
\end{document}